\documentclass{PoS}

\title{New AGN classifications in the Swift/BAT All-Sky Hard X-ray Survey}

\ShortTitle{AGN classifications in the Swift/BAT Survey}

\author{{Pietro Parisi}\\
        INAF-IASF Bologna, via Gobetti 101, 40129 Bologna\\
        Dipartimento di Astronomia, via Ranzani 1, I-40129 Bologna, Italy\\
        E-mail: \email{parisi@iasfbo.inaf.it}}
        
\author{on behalf of the IBIS survey team}

\abstract{Through an optical campaign performed at the San Pedro Martir (Mexico) Telescope and using the 6dF archive ({\tt http://www.aao.gov.au/local/www/6df}, Jones et al. 2004), we determine or give a better classification for 8 newly discovered Active Galactic Nuclei (AGN) in the Swift/BAT 22-months All-sky Hard X-ray Survey (Baumgartner et al. 2008, Tueller et al. 2010). All these objects have observations taken with Swift/XRT or Chandra or XMM archival data which allowed us to pinpoint their optical counterpart thanks to the precise (better than a few arcsec) soft X-ray positions afforded by these observatories.  This information enabled us to obtain optical spectra of all these counterparts, since only three spectra are available on-line, but not flux calibrated, allowing us to reveal their real nature (Baumgartner et al. 2008 give only a tentative classification based upon their X-ray properties). Here we present the spectra, along with the 
corresponding finding charts obtained from the DSS-II red survey, of these 8 sources. We found that our sample is composed of 7 Seyfert 2 and one Seyfert 1, with redshift between 0.009 and 0.068.}

\FullConference{8th INTEGRAL Workshop, `The Restless Gamma-ray Universe' - Integral2010,\\
		September 27-30, 2010\\
		Dublin Ireland}

\begin{document}

\section{Introduction}
The {\it Swift} mission was designed to study cosmic gamma-ray bursts (GRBs) in a multiwavelength context (Gehrels et al. 2004), but with its unique repointing capabilities it is also able to study and monitor other types of X-ray emitting objects.
Through its payload, consisting of three instruments, i.e. the Burst Alert Telescope (BAT; Barthelmy, 2004), the X-Ray Telescope (XRT; Burrows et al. 2004) and the UltraViolet/Optical Telescope (UVOT; Roming et al. 2004), {\it Swift} can detect and follow up X-ray emitting objects over a wide range of wavelengths.

In this work we have classified 8 AGNs either without optical identification,
or without published optical spectra, belonging to the {\it Swift}/BAT AGN surveys of Tueller et al. (2010), with a soft X-ray position and a tentative classification based upon their X-ray properties given by Baumgartner et al. (2008) (except for the three sources in the 6dF archive).  
Following the method applied by Masetti et al. (2004, 2006a,b, 2008, 2009, 2010) for the optical spectroscopic follow-up of unidentified {\it INTEGRAL} sources, we confirm the AGN nature and give the exact classification of these 8 objects.

\section{Observations and data analysis}
The optical spectroscopic data presented here (see Tab. \ref{res}) were obtained with the 2.1m telescope of the Observatorio Astr\'onomico Nacional in San Pedro Martir, M\'exico; we also used archival spectra from the 6dF archive \footnote{{\tt http://www.aao.gov.au/local/www/6dF}} (Jones et al. 2004).

The SPM data reduction was performed with the standard procedure (optimal extraction; Horne 1986) using IRAF\footnote{
IRAF is the Image Reduction and Analysis 
Facility made available to the astronomical community by the National 
Optical Astronomy Observatories, which are operated by AURA, Inc., under 
contract with the U.S. National Science Foundation. It is available at 
{\tt http://iraf.noao.edu/}}.
Calibration frames (flat fields and bias) were taken on the day preceding or following 
the observing night. The wavelength calibration was obtained using lamp spectra 
acquired soon after each on-target spectroscopic acquisition. The uncertainty on the 
calibration was $\sim$0.5~\AA~for all cases. This was checked using the positions of 
background night sky lines. Flux calibration was performed using 
catalogued spectrophotometric standards.
Since the 6dFGS archive provides spectra which are not calibrated in flux, we used the optical photometric information in 
Jones et al. (2005) and Doyle et al. (2005) to calibrate the 6dFGS data presented here.
Objects with more than one observation had their spectra stacked 
together to increase the signal-to-noise ratio.

\section{The sample}
The identification and classification approach we adopt in the analysis of the optical spectra is the following:
for the emission-line AGN classification, we used the criteria 
of Veilleux \& Osterbrock (1987) and the line ratio diagnostics 
of Ho et al. (1993, 1997) and of Kauffmann et al. (2003); for 
the subclass assignation to Seyfert 1 galaxies, we used the 
\mbox{H$_\beta$/[O {\sc iii}]$\lambda$5007} line flux ratio criterion as in
Winkler et al. (1992).

\begin{figure*}[th!]
\begin{center}
\includegraphics[width=12cm,angle=0]{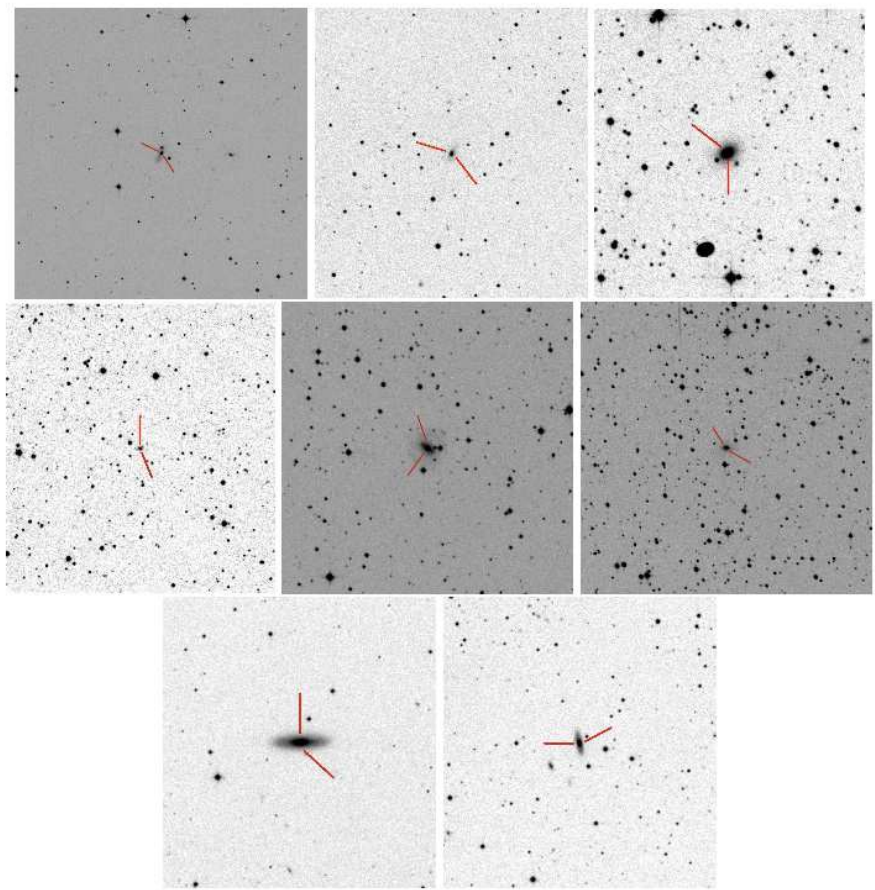}
\caption{From left to right and top to bottom: optical images of the
fields of Swift J0100.9$-$4750, Swift J0248.7+2626, Swift J0543.7$-$2741, Swift J0544.3+5910, Swift J0623.8$-$3215,
Swift J0923.9$-$3143, Swift J1246.9+5433 and Swift J2341.8+3034.
The optical counterparts of the {\it Swift} sources are indicated with tick marks. Field sizes are
5$'$$\times$5$'$ and are extracted  from the DSS-II-Red survey. In all
cases, North is up and East to the left.}
\label{find}
\end{center}
\end{figure*}

\begin{figure*}[th!]
\begin{center}
\includegraphics[width=4cm,angle=270]{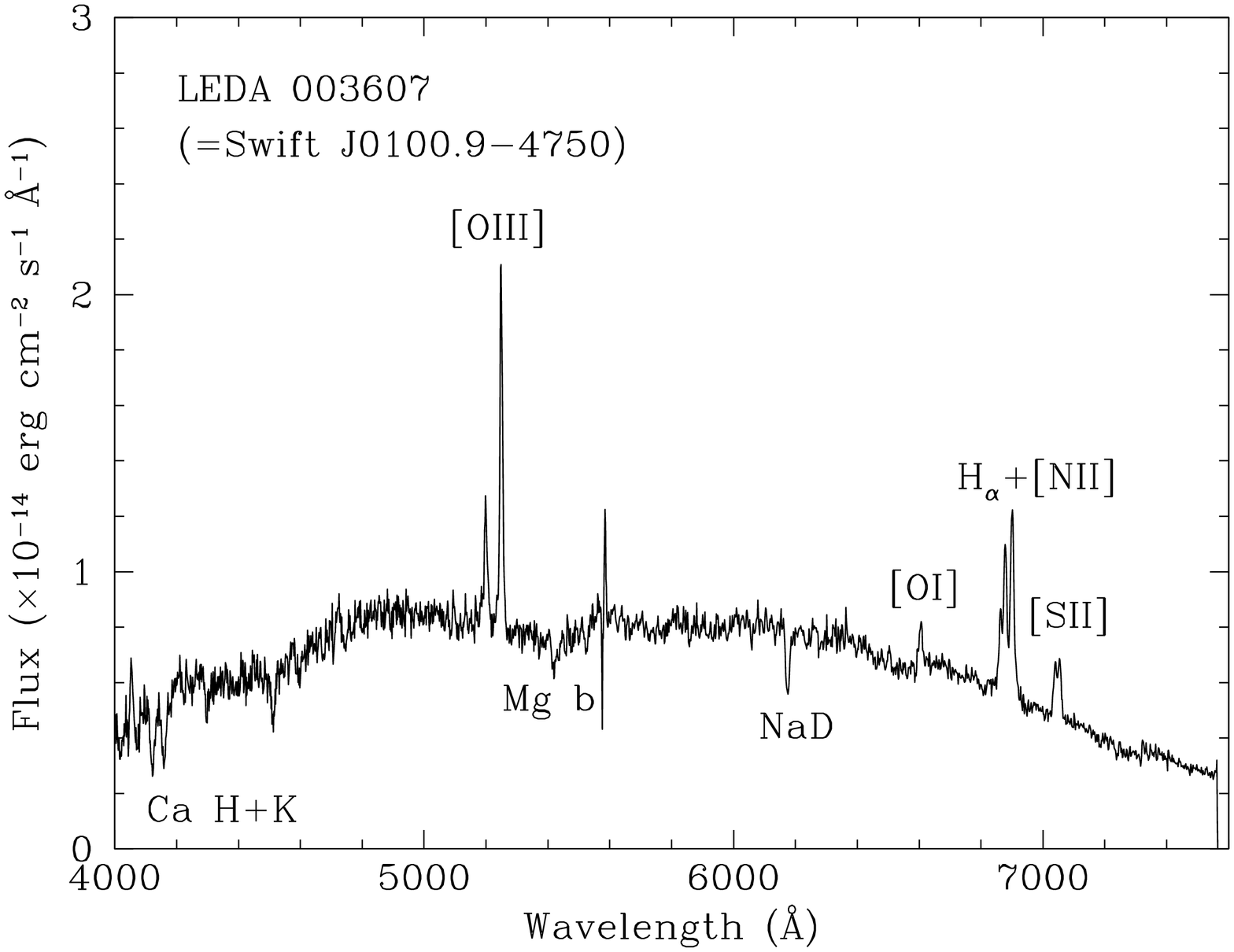}
\includegraphics[width=4cm,angle=270]{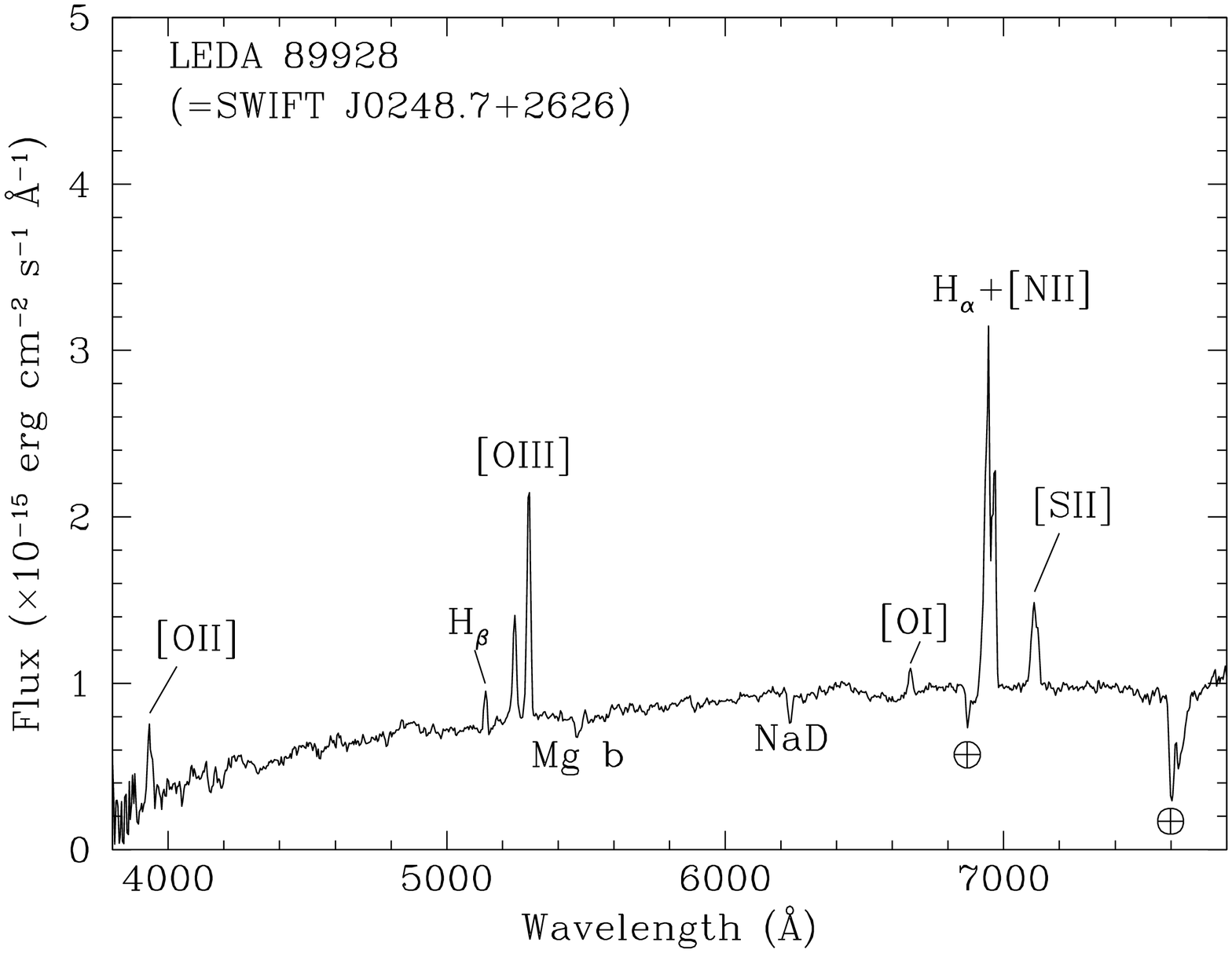}
\includegraphics[width=4cm,angle=270]{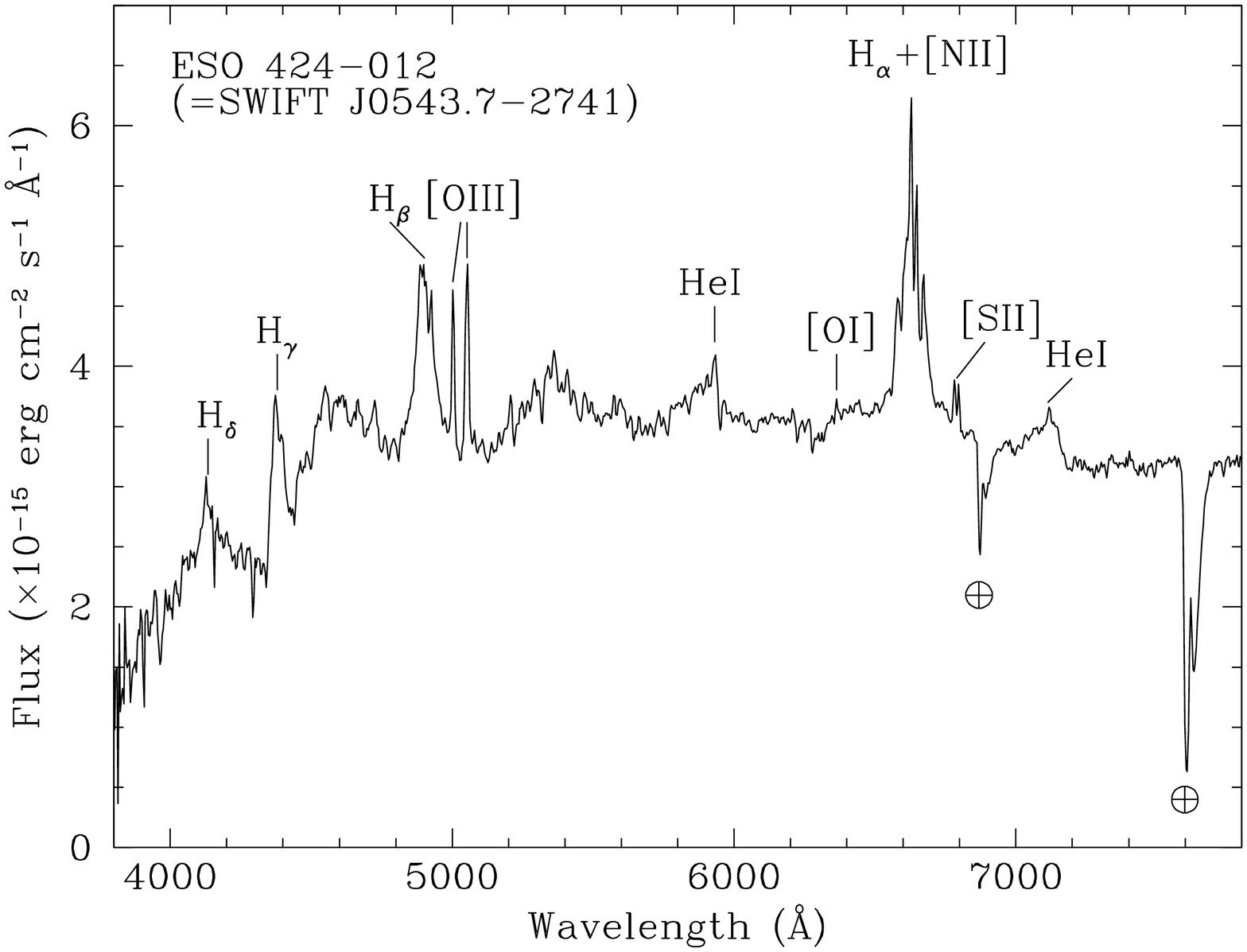}
\includegraphics[width=4cm,angle=270]{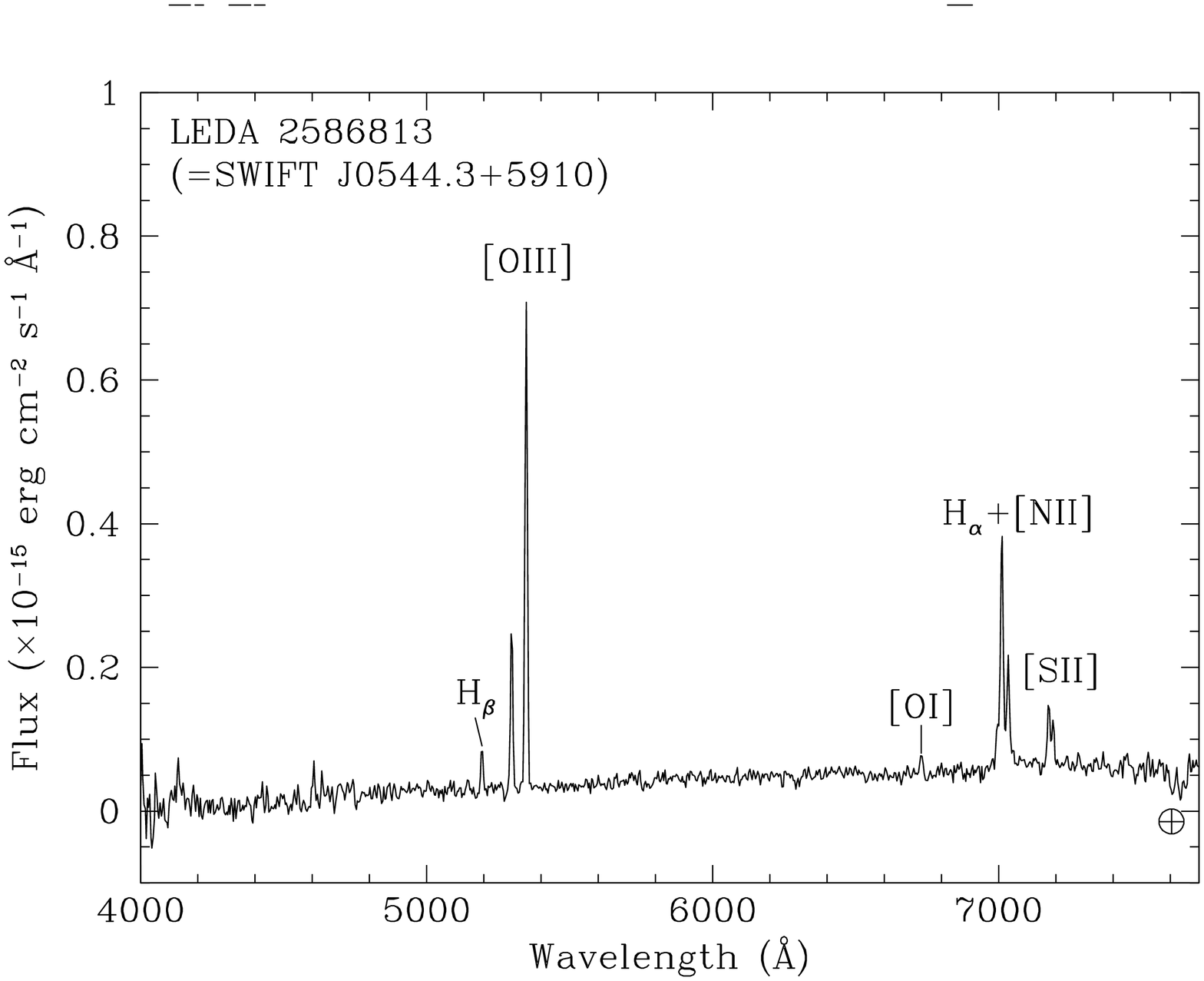}
\includegraphics[width=4cm,angle=270]{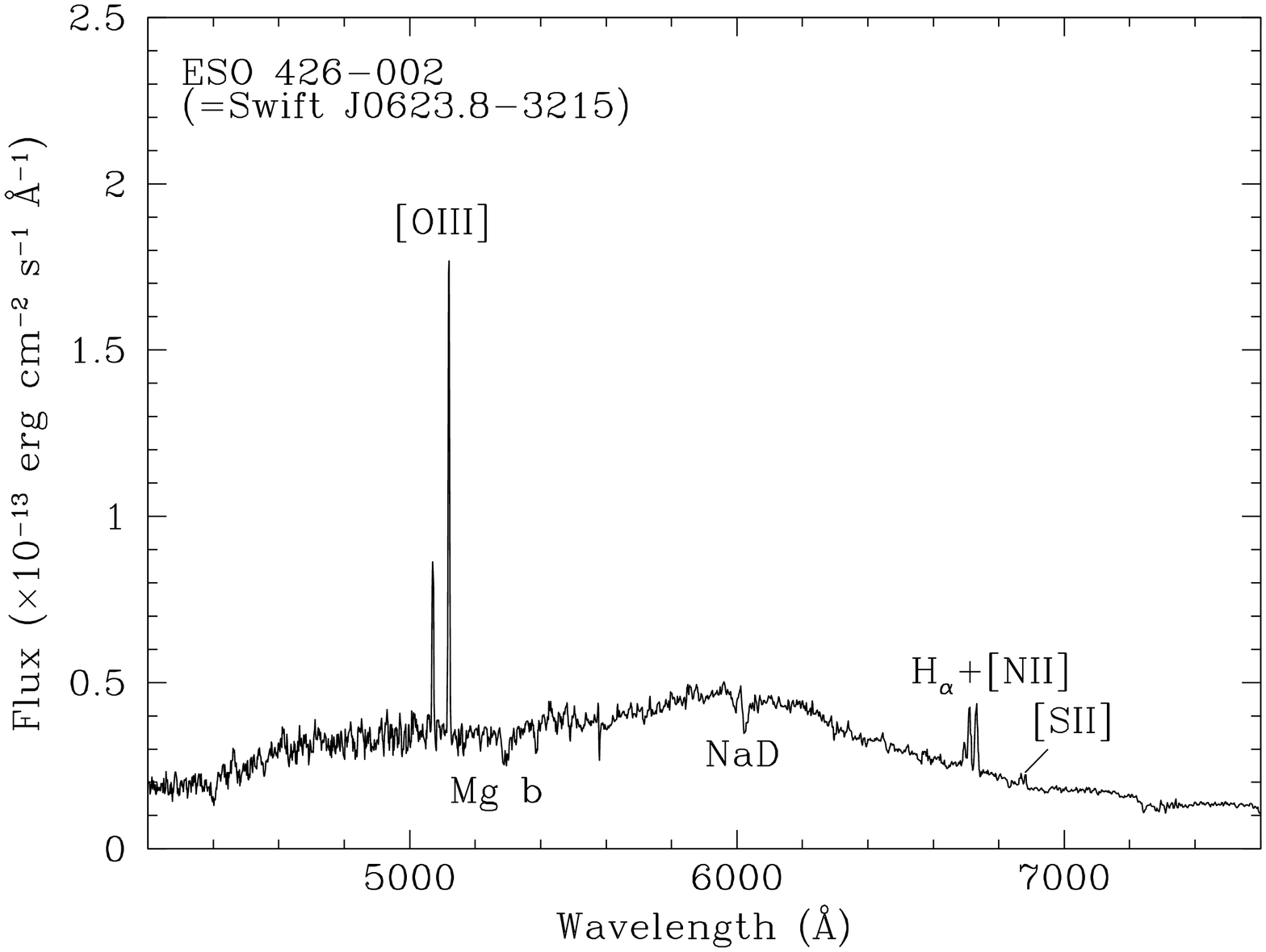}
\includegraphics[width=4cm,angle=270]{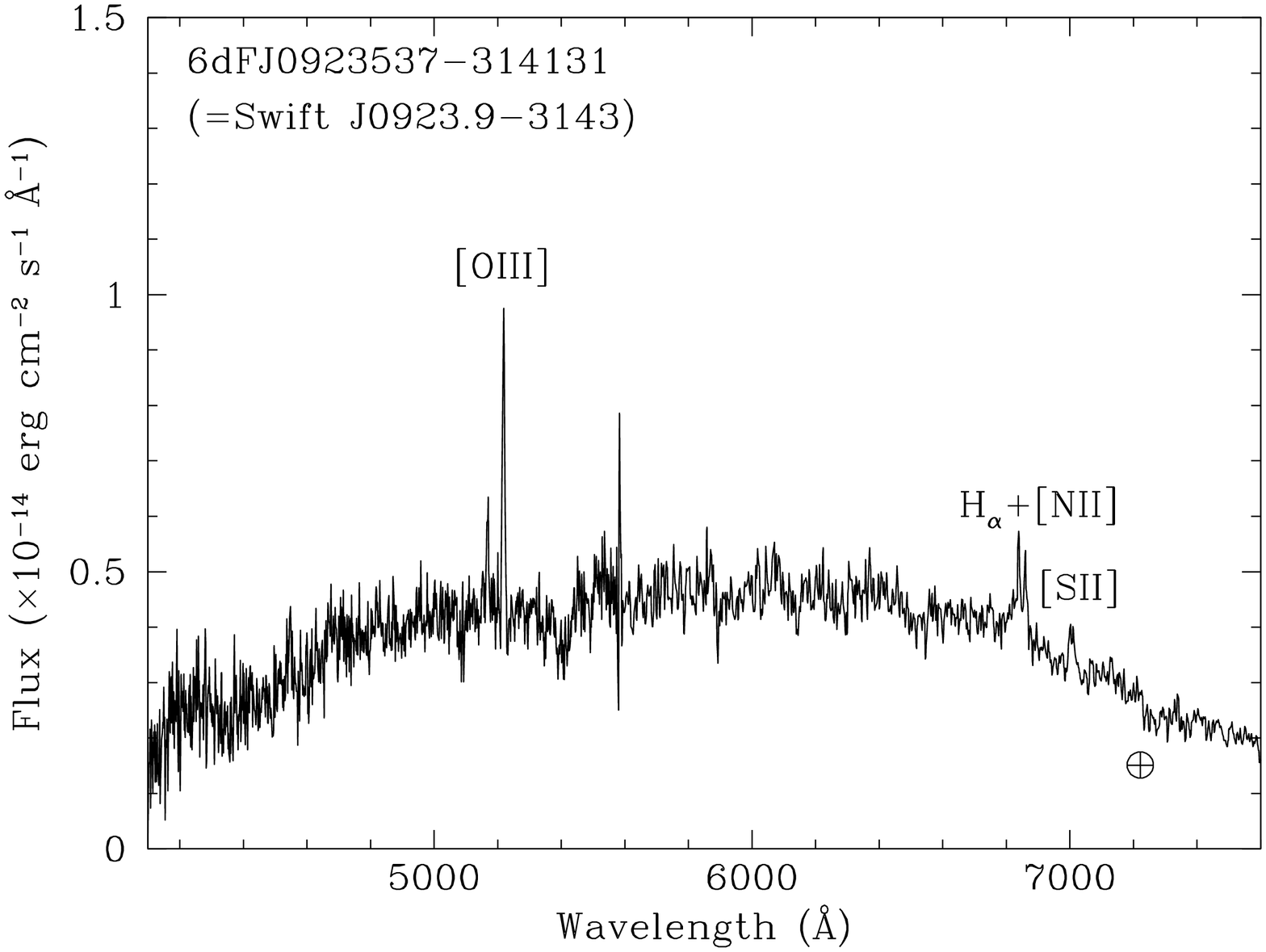}
\includegraphics[width=4cm,angle=270]{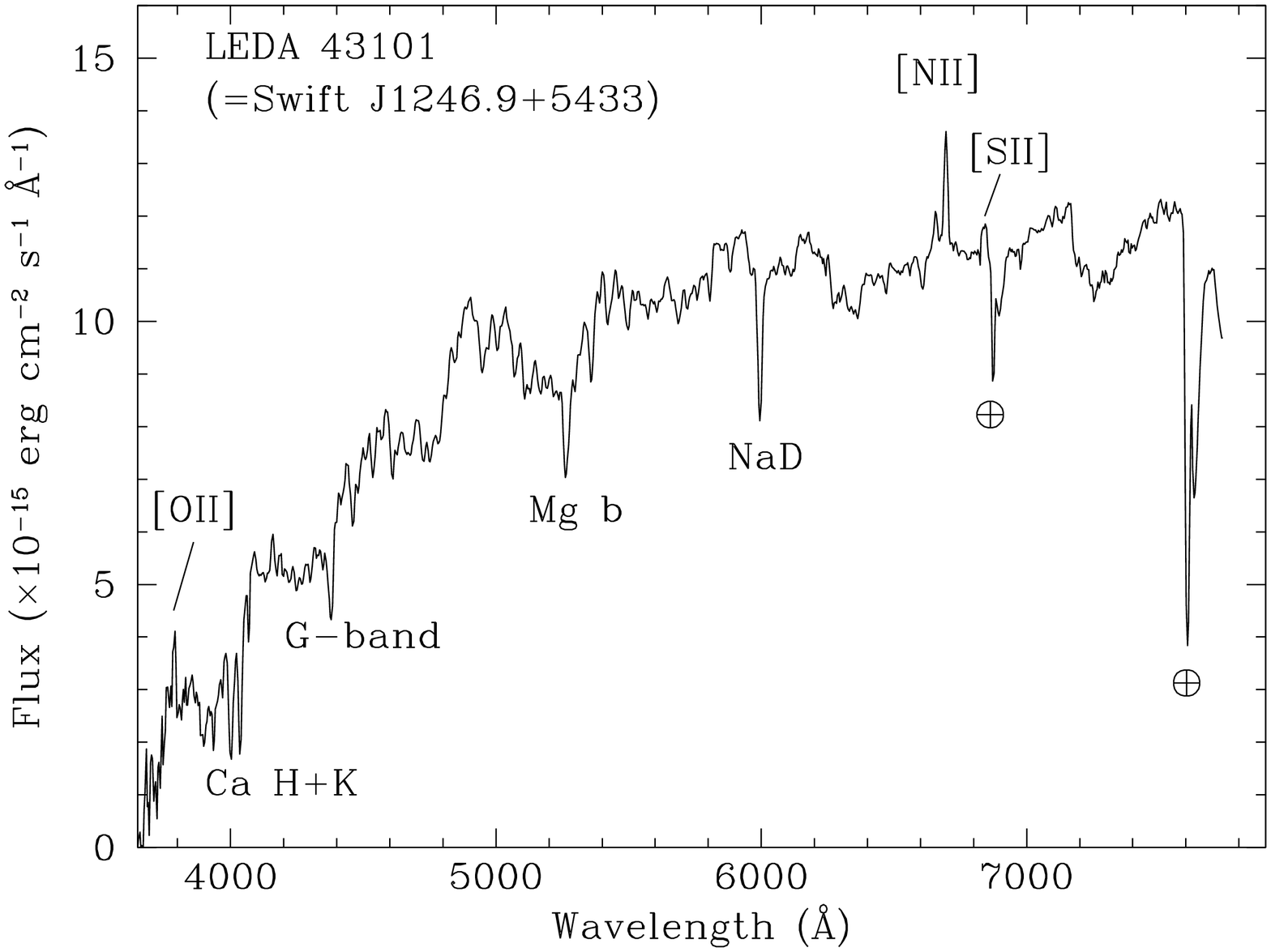}
\includegraphics[width=4cm,angle=270]{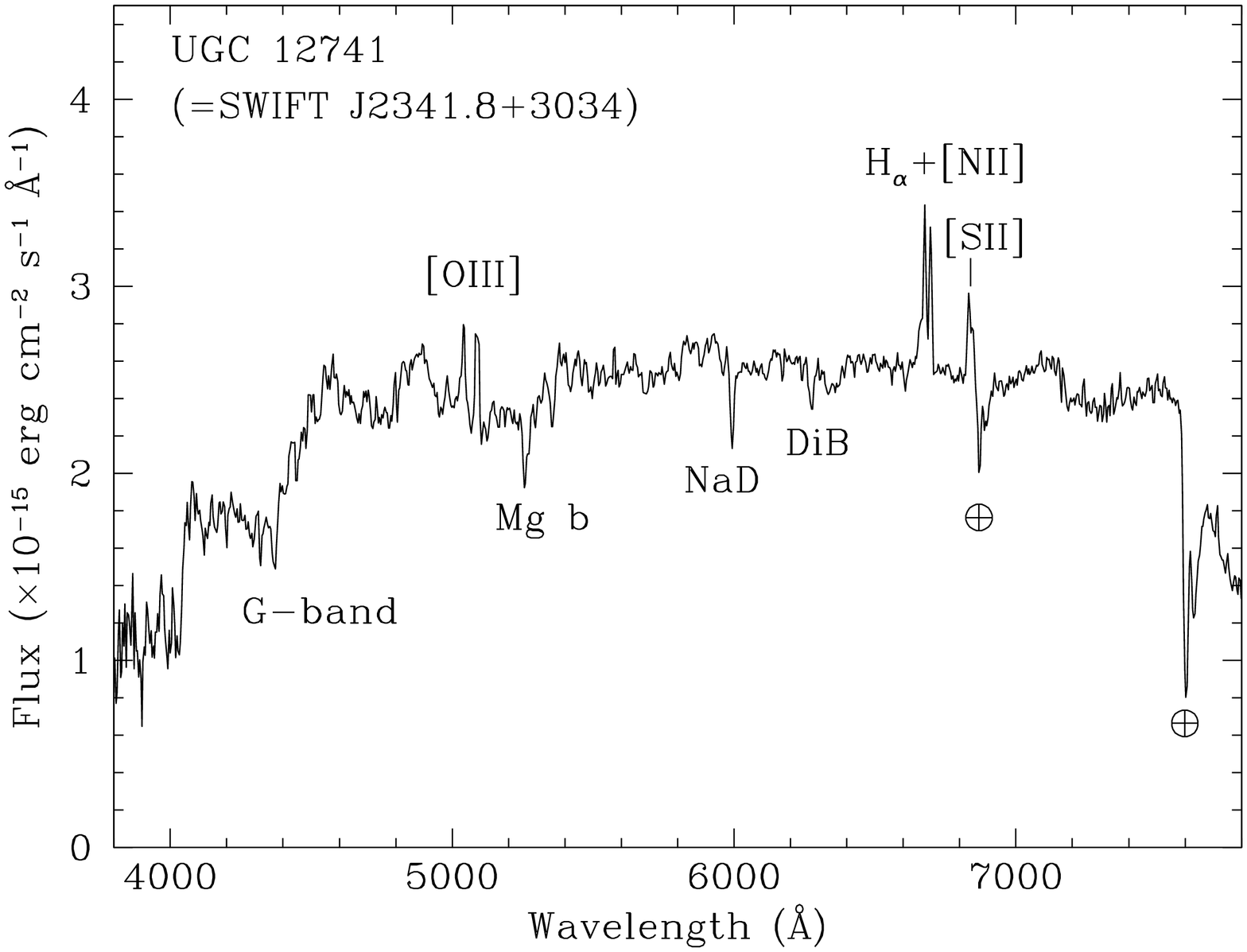}
\caption{Spectra (not corrected for the intervening Galactic absorption) of  the 
optical counterpart of AGNs Swift J0100.9$-$4750, Swift J0248.7+2626, Swift J0543.7$-$2741, Swift J0544.3+5910, Swift J0623.8$-$3215,
Swift J0923.9$-$3143, Swift J1246.9+5433 and Swift J2341.8+3034.}
\label{spectra}
\end{center}
\end{figure*}

Following these criteria we found 7 type 2 AGNs and one type 1 AGN (see Fig.\ref{spectra}).
For each source we reported the finding charts (Fig.\ref{find}).
Concerning the type 2 AGN spectra, we found that for 5 sources the continuum is quite regular 
(Swift J0100.9$-$4750, Swift J0248.7+2626, Swift J0544.3+5910, Swift J0623.8$-$3215,
 Swift J0923.9$-$3143), while for the remaining objects
(Swift J1246.9+5433 and Swift J2341.8+3034) the continuum is dominated by the star content of the host galaxy. 
In most of the spectra the [O {\sc iii}]$\lambda$5007 and [N{\sc ii}]$\lambda$6585 forbidden lines, and the Balmer 
H$_\alpha$ permitted line are well detected in emission. The spectra are not corrected for the Galactic extinction. 
We determine that their redshifts lie between 0.017 and 0.068.

The single type 1 AGN (Swift J0543.7$-$2741) shows broad H$_{\alpha}$+[N{\sc ii}] complex in emission, 
the Balmer series up to H$_{\delta}$ in emission as well, the [O{\sc iii}] and [S{\sc ii}] forbidden narrow emission lines, and He{\sc i} broad emissions.
The narrow lines enabled us to calculate the redshift of this object, that is 0.009.
The continuum is not regular, contaminated by the star content of the underlying galaxy.

\begin{table*}[th!]
\begin{center}
\caption[]{Main results obtained from the analysis of the optical spectra of the 8 AGNs of the present 
sample}\label{res}
\scriptsize
\resizebox{15cm}{!}{
\begin{tabular}{lccccc}
\noalign{\smallskip}
\hline
\hline
\noalign{\smallskip}
\multicolumn{1}{c}{Object} & Opt. counterpart &Class & $z$  & Opt. telescope or archive \\
\noalign{\smallskip}
\noalign{\smallskip}
\hline
\noalign{\smallskip}
Swift J0100.9$-$4750 &2MASX J01003490-4752033 & Seyfert 2 & 0.048 & AAT+6dF \\
Swift J0248.7+2626& 2MASX J02485937+2630391&Seyfert 2 & 0.057 & San Pedro Martir \\
Swift J0543.7$-$2741& MCG -05-14-012&Seyfert 1.2 & 0.009 & San Pedro Martir & \\
Swift J0544.3+5910 &2MASX J05442257+5907361 &Seyfert 2 & 0.068 & San Pedro Martir & \\
Swift J0623.8$-$3215&ESO 426-G 002 & Seyfert 2 & 0.022 &AAT+6dF \\
Swift J0923.9$-$3143&2MASX J09235371-3141305 &Seyfert 2 & 0.042 & AAT+6dF \\
Swift J1246.9+5433 &NGC 4686 &Seyfert 2& 0.017 & San Pedro Martir  \\
Swift J2341.8+3034 & UGC 12741&Seyfert 2 & 0.017& San Pedro Martir   \\
\noalign{\smallskip} 
\hline
\noalign{\smallskip}  
\multicolumn{5}{l}{The typical error on the redshift measurement is $\pm$0.001 
but for 6dFGS spectra we can assume an uncertainty} \\
\multicolumn{5}{l}{of $\pm$0.0003.} \\
\noalign{\smallskip} 
\hline
\hline
\end{tabular}}
\end{center}
\end{table*}

\newpage
\section{Conclusion}
In this work we have either given for the first time (Swift J0248.7+2626, Swift J0543.7$-$2741, Swift J0544.3+5910, Swift J1246.9+5433, Swift J2341.8+3034), 
or presented a better classification (Swift J0100.9$-$4750, Swift J0623.8$-$3215, Swift J0923.9$-$3143) of, 
the optical spectroscopic identification for 8 Swift AGNs belonging to the Swift/BAT 22-months All-sky Hard X-ray Survey. 
This was achieved through an observational campaign in Mexico, at the San Pedro Martir Telescope and with the use of the 6dF archive. 
We found that our sample is composed of 8 AGNs (7 Seyfert 2 and one Seyfert 1), with redshift between 0.009 and 0.068.


\end{document}